\documentclass[article,titlepage,a4paper,reprint,superscriptaddress, floatfix]{revtex4-1}

\usepackage[dvipsnames]{xcolor}
\usepackage{amsfonts}
\usepackage{physics}
\usepackage{hyperref}  
\usepackage{mathtools}
\usepackage{listings}
\usepackage{bm}
\usepackage{latexsym}
\usepackage{times} 

\usepackage{graphicx} 
\graphicspath{{figures/}} 
\usepackage{booktabs} 
\usepackage{amsfonts, amsmath, amsthm, amssymb} 
\usepackage{graphicx}
\usepackage{bm}

\usepackage{soul} 

\begin{document}


\title{Flat-band-induced superconductivity in synthetic bilayer optical lattices }

\author{Tymoteusz Salamon\thanks{tymoteusz.salamon@icfo.eu}}
\affiliation{ICFO - Institut de Ciencies Fotoniques, The Barcelona Institute of Science and Technology, Av. Carl Friedrich Gauss 3, 08860 Castelldefels (Barcelona), Spain}
\author{Bernhard Irsigler}
\affiliation{ICFO - Institut de Ciencies Fotoniques, The Barcelona Institute of Science and Technology, Av. Carl Friedrich Gauss 3, 08860 Castelldefels (Barcelona), Spain}
\author{Debraj Rakshit}
\affiliation{Harish-Chandra Research Institute, A CI of Homi Bhabha National Institute, Chhatnag Road, Jhunsi, Allahabad 211 019, India}
\author{Maciej Lewenstein}
\affiliation{ICFO - Institut de Ciencies Fotoniques, The Barcelona Institute of Science and Technology, Av. Carl Friedrich Gauss 3, 08860 Castelldefels (Barcelona), Spain}
\affiliation{ICREA, Pg. Lluis Companys 23, Barcelona, Spain}

\author{Tobias Grass}
\affiliation{ICFO - Institut de Ciencies Fotoniques, The Barcelona Institute of Science and Technology, Av. Carl Friedrich Gauss 3, 08860 Castelldefels (Barcelona), Spain}
\affiliation{DIPC - Donostia International Physics Center, Paseo Manuel de Lardiz{\'a}bal 4, 20018 San Sebasti{\'a}n, Spain}
\affiliation{Ikerbasque - Basque Foundation for Science, Maria Diaz de Haro 3, 48013 Bilbao, Spain}
\author{Ravindra Chhajlany}
\affiliation{Faculty of Physics, Adam Mickiewicz University, 61614 Pozna\'n, Poland}
\date{\today}

\begin{abstract}
 Stacking two layers of graphene with a relative twist angle gives rise to moir\'e patterns, which can strongly modify electronic behavior  and may lead to unconventional superconductivity. A synthetic version of twisted bilayers can be engineered with cold atoms in optical lattices. Here, the bilayer structure is mimicked through  coupling between atomic sublevels, and the twist is achieved by a spatial modulation of this coupling. In the present paper, we investigate the superconducting behavior of fermionic atoms in such a synthetic twisted bilayer lattice. Attractive interactions between the atoms are treated on the mean-field level, and the superconducting behavior is analyzed via the self-consistently determined pairing gap. A strong enhancement of the pairing gap is found, when a  quasi-flat band structure occurs at the Fermi surface, reflecting the prominent role played by the twist on the superconductivity. The tunability of interactions allows for the switching of superconducting correlations from intra (synthetic) layer to inter (synthetic) layer. This includes also the intermediate scenario, in which the competition between inter- and intra-layer coupling completely destroys the superconducting behavior, resulting in re-entrant superconductivity upon tuning of the interactions.
\end{abstract}
\maketitle
\section{\label{intro}Introduction}
The Fermi-Hubbard model plays a central role in describing various aspects of the many-body physics of condensed matter systems
\cite{Hensgens_2017,Esslinger10,bloch08,Lewenstein07,Truscott01,Chen21}. In particular, along with its variants, it is 
 widely believed to  encompass the basic ingredients required to understand  high-temperature superconductivity, \textit{e.g.} in  cuprates  \cite{Micnas-rmp, Arovas-review,Feiner96,berg08,Macridin_2005}. 
 Although Hubbard models are effective simplified models of complex condensed matter systems, they can be realized with high fidelity and control in various engineered systems  
such as ultracold atoms in optical lattices. This, in turn, has lead to an exciting branch of physics - quantum simulation of condensed matter phenomena \cite{Lewenstein12,Bloch12,GRYNBERG01,Hofstetter02}. In recent years, one-, two- and three-dimensional optical lattices have been generated paving the way to  studies of different quantum phases under various types of interactions \cite{Mazurenko2017, Schenider08}. A distinguishing feature within the cold atom set-up is the freedom of precise tuning of microscopic system parameters, {\it  i.e.} interactions and particle tunneling over wide ranges  \cite{Chin10}.

The control of material properties via band structure engineering has been a  long standing goal in condensed matter physics.  A new frontier is twistronics, where the relative rotational misalignment between layers in quasi two dimensional (2D) systems leads to moir\'e patterns in real space. The moir\'e patterns strongly influence the band structure and lead to enhanced collective effects induced by interactions and topology \cite{Andrei2021}.  In graphene bilayers,  tuning the twist angle to so-called magic values was predicted to  strongly quench the electronic kinetic energy leading to the formation of quasi-flat bands   \cite{CastroNeto2007,bistritzer11,McCann_2013}, where   small interactions can   dominate the phenomenology~\cite{Torma2022_review}.  With the successful development of fabrication methods for such devices,  a series of experiments spectacularly unveiled superconductivity and correlated insulators in these materials around the magic angles~\cite{Cao2018,Cao18_2,Wang2020}. The enhancement of superconductivity  originating from completely isolated or non-isolated quasi-flat bands, in particular, has turned out to be an exciting development in the search for high-Tc superconductivity driven by quantum geometry ~\cite{julku16,Huhtinen22,Hu19,Heikkil_2011}.

These results have driven the new field of twistronics involving the study of various kinds of Van der Waals stacked heterostructures beyond bilayer graphene \cite{Kennes2021a}.   The unavoidable effect of twisting 2D materials   is the enlargement of the unit cell,  usually by a few orders of magnitude compared to the original unit cell of a single sheet of the material \cite{Andrei2021,kim17},  into a so-called moir\'e supercell.  This emergent approximate crystal symmetry strongly complicates direct microscopic modelling and non-approximating studies of correlations in these systems.  As a result direct quantum simulations of twistronics,  in particular based on the promising platform of ultra cold atoms trapped in optical lattices,  that allow for exquisite control of system parameters,  offer an additional  window to gain fundamental understanding of moir\'e materials.   Importantly,  this approach allows to study systems without certain practical difficulties associated with materials  such as the lack of control over the homogeneity of the twist angle in  samples and strain effects which lead to disorder.  Moreover,  apart from the control of interactions,  ultra cold atom systems allow for tuning of interlayer coupling to strong values  that can lead to enhanced correlation effects even for comparatively small moir\'e supercells.

The interest in general 2D bilayer systems has led to the design of multiple architectures and control schemes for bi-layer optical lattices \cite{grass16,Kantian2018,Gall2021}. Moir\'e systems  can be generated, on one hand,  by effectively performing  the direct analog of material twisting in overlapping samples \cite{Tudela2019}, or  spin dependent optical lattices  \cite{Meng21,Luo21}.  A different versatile approach stems from the fact that physically the main effect of twisting is the induction of incommensurate quasi-periodic potentials and quasi-periodic interlayer tunnelings in layered systems \cite{Macdonald2018, Vishwanath2019,Fu2020,Chou2020,Salamon_2020,Lee2021,Yi2022}.  Such spatially modulated patterns can be directly imprinted on synthetic bilayer systems \cite{Salamon_2020,Sala2},  {\it i.e.} 
a single physical optical lattice layer of atomic species  with  Raman coupled  internal states  playing the role of the additional layer degree of freedom ~\cite{Boada_2012,celi14}.  This remarkably realizes moir\'e-type physics without physical twisting. 
\\
\\
In the present paper, we build on the idea originally presented in \cite{Salamon_2020} which uses the concept of synthetic dimensions to engineer twisted bilayers. 
For such scenario, we consider attractive on-site $s$-wave interactions with full SU(4) symmetry and explore superconducting properties in such synthetic bi-layers with a chosen size and shape of the supercell. 
Near-flat bands with very small dispersion compared to its immediately neighboring bands can be accessed with rather small unit cells in our synthetic-dimension-based proposal, which allows us to adopt an multi-band Hartree-Fock-Bogoliubov theory \cite{osti79,lewin14} for probing superconductivity. The analysis is performed extensively for a wide range of experimentally controllable parameters, such as interlayer coupling and interaction strength. Our study seeks to understand and to characterize the role played by a finite dispersion of the quasi-flat band on superconductivity.  In fact, the proposed setup allows to accurately control the widths of the quasi-flat bands over a broad range \cite{Salamon_2020,Salamon_2021}. 

The paper is arranged as follows: In Sec.~\ref{system} we present the lattice Hamiltonian and discuss the band structure and interaction types appearing in the model. In Sec.~\ref{MF} the detailed description of Hartree-Fock-Bogoliubov mean field decoupling is shown together with resulting Bogoliubov - de Gennes Hamiltonian and self-consistent procedure. Sec.~\ref{results} presents original results of our study. First, we take into account  the impact of band flattening, caused by modulation of the inter-layer hopping, on the superconducting gap. Then, we consider breaking the symmetry into SU(2) $\times$ SU(2) via selectively tuning the interaction channels. We also tune one of interaction types to be of negligible amplitude  to facilitate the comparison of the obtained results with standard bi-layer Fermi-Hubbard system. Finally, conclusions are presented in Sec.~\ref{Conclusions}.

\section{\label{system}The system}
\begin{figure*}
\makebox[\textwidth][c]{\includegraphics[width=1\textwidth]{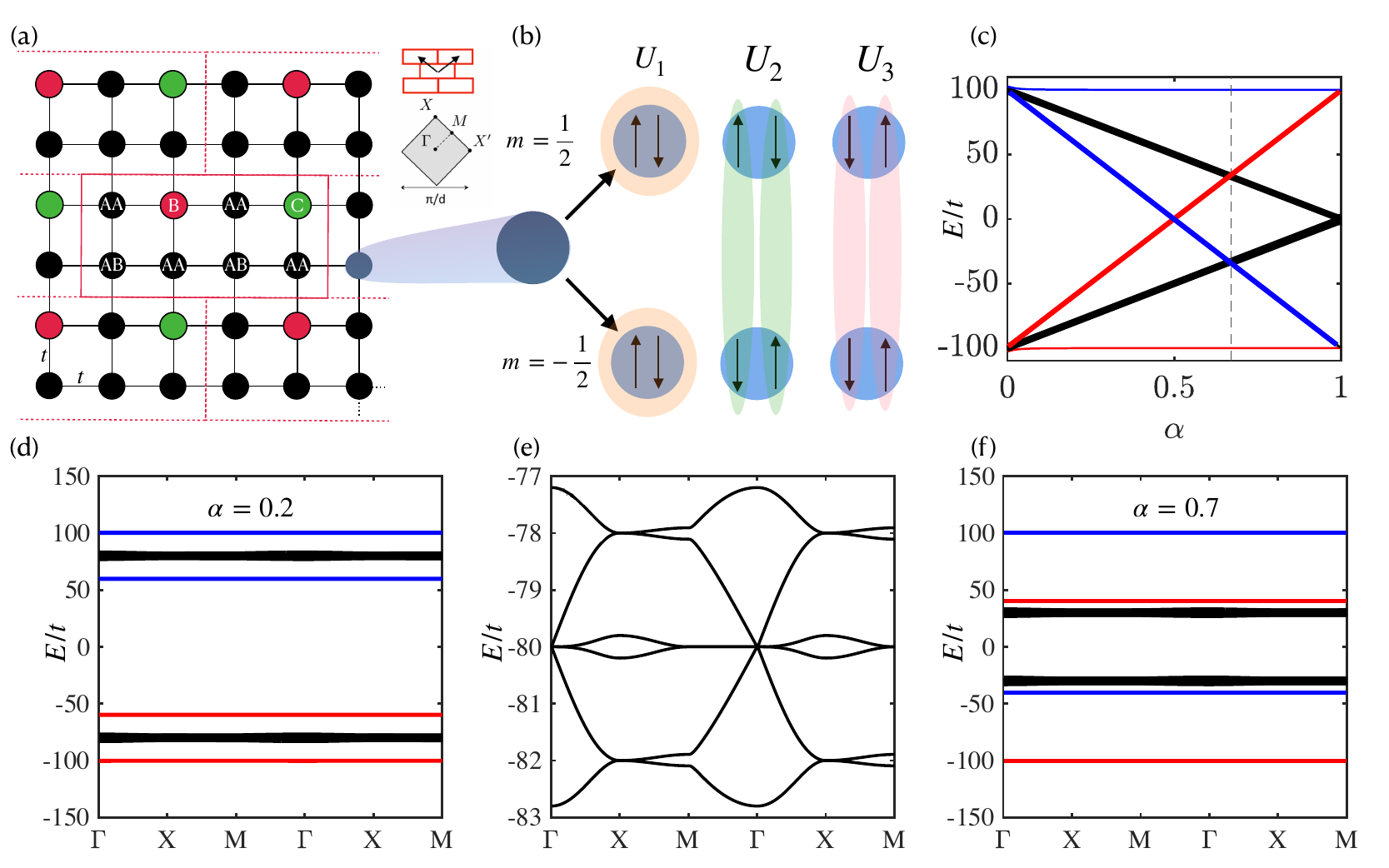}}%
\caption{
(a) Square lattice with unit cell of $8$ physical sites due to synthetic coupling which is different on the black ($AA$ and $AB$ - depending on the amount of A-type neighbours), green ($B$), and red ($C$) sites. The lattice is covered by unit cells in a brick wall arrangement, as indicated in the inset. 
(b) Schematic representation of the three interaction types appearing in the system.
(c) Evolution of the average energy of the bands as a function of the modulation parameter $\alpha$ at $\Omega_0=100$ (in units of $t$). The panel shows the outer bands (two blue lines and two read ones) of each, positive and negative manifold being shifted with respect to two the inner subsets of bands marked with black lines. At the critical value of $\alpha_c\approx 0.67$ the blue band from the upper manifold reaches lower energy than black, negative manifold. This results with change of the position of the Fermi surface.
(d) Full spectrum of the system under  periodic modulation $\Omega_0=100$ and $\alpha=0.2$. 
(e) Zoomed plot of (d) highlighting the negative six-fold subset of bands including the central quasi-flat bands. 
(f) Spectrum of the system at $\Omega_0=100$ and $\alpha=0.7$, after the flipping of the bands occurred (dashed line in panel (c)).}
\label{main_fig}
\end{figure*}

The considered cold-atom system consists of $N$ fermionic atoms loaded into an optical square lattice with unit lattice constant. Atoms with large nucleus manifold, e.g., $^{87}$Sr or $^{173}$Yb \cite{cazalilla14}, are prepared in four internal states, which are then interpreted as two spin levels within two synthetic layer levels. Accordingly, the four levels are denoted by two quantum numbers, $\sigma=\uparrow,\downarrow$ for the (pseudo-)spin, and ${m=+,-}$ for the (pseudo-)layer. Throughout this study, the hopping amplitude $t$ is set to unity, i.e. $t=1$, which fixes the units of the energies reported in this work. The synthetic hopping amplitude between the pseudo-layers, i.e., between $m=+$ and $m=-$, is denoted by $\Omega({\bf r})$. The synthetic hopping is generated through a pair of Raman lasers, and the position dependence of this coupling can be used to produce a synthetic twist, which gives rise to the flattening of certain energy bands, see \cite{Salamon_2020}. Accordingly, the kinetic part of the Hamiltonian reads
\begin{equation}
\label{eq:Ht}
H_{\rm kin} = H_t+H_\Omega, 
\end{equation}
where 
\begin{equation}
\begin{split}
H_t =& -t\sum_{{\bf r},m,\sigma}  \left[a_{m,\sigma}^{\dagger}({\bf r}+\mathbf{1}_x) + a_{m,\sigma}^{\dagger}({\bf r}+\mathbf{1}_y) \right]a_{m,\sigma}({\bf r})\\
&+\mathrm{h.c.}
\end{split}
\end{equation}
denotes the hopping Hamiltonian between different sites in the optical lattice, and
\begin{equation}
\label{homega}
H_\Omega = \sum_{{\bf r},m,m',\sigma} \Omega({\bf r})~a_{m,\sigma}^{\dagger}({\bf r})a_{m',\sigma}({\bf r}) +\mathrm{h.c.}
\end{equation}
 denotes the synthetic hopping Hamiltonian. Here, $a_{m,\sigma}^\dag(\bf{r})$ and $a_{m,\sigma}(\bf{r})$ represent the fermionic creation and annihilation operators, respectively and  ${\bf r} = (x,y)$ denotes the position of a lattice site in the two-dimensional plane of the optical square lattice, where $x$ and $y$ are integers. Moreover, we define the unit vectors $\mathbf{1}_x=(1,0)$ and $\mathbf{1}_y=(0,1)$. Note that in Eq.~\eqref{homega} we have chosen a spatially constant phase of the coupling, but it is also straightforwardly possible to implement a position-dependent phase term. Such a choice would allow for incorporating artificial gauge fields into the synthetic dimension \cite{celi14,price17}, and the effect of such artificial gauge fields in the context twist-simulating optical lattices has been already discussed in \cite{Salamon_2020}. 
 The spatial modulation of the synthetic hopping strength is given by
\begin{equation}
\label{omega_form}
     \Omega({\textbf{r}}) = \Omega_0 \left\{1 -\alpha[1+\cos{(2 \pi x/l_x)} \cos{(2 \pi y/l_y)}] \right\}.
\end{equation}
The strength of the spatial modulation is controlled by the dimensionless parameter $\alpha$ which imposes a twist on the hopping energies in the lattice, i.e., a Moir\'{e} lattice. The two length scales, $l_x$ and $l_y$, define the size of the unit cell of the Moir\'{e} lattice. Here, we focus on $l_x=l_y=4$, which is the simplest case of the, so-called, first "magic-configuration". This choice is, yet, large enough to make the band structure sufficiently flat, see Ref.~\cite{Salamon_2020}, but still keeps the unit cell small enough for an efficient computational treatment. It is important to note that, despite the original lattice being a square lattice, the Moir\'{e} lattice has a graphene-like brick-wall geometry. Its unit cell contains $8$ physical sites of the original lattice, as shown in Fig.~\ref{main_fig}(a).Among these eight sites, we distinguish four sets of sites $\{AA,AB,B,C\}$:  $AA$ and $AB$ denote the six black sites in Fig.~\ref{main_fig}(a) with two or four "black" nearest neighbours, respectively , and $B$ and $C$ denote the red and green sites, respectively. According to Eq.~\eqref{omega_form}, these three sets of sites exhibit the following property:

\begin{equation}
\Omega(\bf{r})=
    \begin{cases}
    \Omega_0(1-\alpha)&\text{ if }{\bf r}\in AA \lor AB,\\
    \Omega_0&\text{ if }{\bf r}\in B,\\
    \Omega_0(1-2\alpha)&\text{ if }{\bf r}\in C,\\
    \end{cases}
    \label{eq_unitcell}
\end{equation}
that is, the coupling strength at sites B and C is shifted by $\abs{\alpha}$ with respect to the coupling strength at sites of type A, that is, type AA or AB. 
\\

For our choice of $l_x$ and $l_y$, the spectrum of $H_{\rm kin}$ consists of 16 energy bands. A symmetrical arrangement of the bands with respect to $E=0$ reflects the particle-hole symmetry in the system. At sufficiently large interlayer tunneling, e.g., $\Omega_0=100t$,the spectrum is gapped, as shown on Fig.~\ref{main_fig}(d) for $\alpha=0.2$ and (f) for $\alpha=0.7$. Among these 16 bands, we focus on the subset of six bands at $E\approx\pm80t$, i.e. we assume a Fermi energy in the vicinity of this manifold. These two manifolds are plotted in black in Fig.~\ref{main_fig}(d), and further analyzed in Fig.~\ref{main_fig}(e), where we zoom into one of the six-fold manifold. Importantly, we observe that the two central bands of this manifold are almost flat. Moreover, there are different degenercies, e.g. at the $\Gamma$-point. In addition to these quasi-flat bands, the system also exhibits isolated flat bands shown in blue and red in Fig.~\ref{main_fig}(d). 

The impact of the twist parameter $\alpha$ on the structure of the whole energy spectrum is shown in Fig.~\ref{main_fig}(b). For $\alpha<0.67$, the isolated bands are located above and below the six-fold manifolds, whereas for $\alpha>0.67$ both isolated bands are above the sixfold manifold on the positive side of the spectrum, and below the sixfold manifold on the negative side of the spectrum. 

Within the sixfold manifold, the modulation of the interlayer tunneling flattens the two central bands, as it mimics Moir\'{e} patterning. This implies that, for $\alpha<0.67$, this quasi-flat band structure appears at the Fermi surface, when the filling $\nu$ of the system is 1/4 or 3/4, i.e., the lowest four or the lowest twelve bands are filled, see Fig.~\ref{main_fig}(d) and (e). For $\alpha>0.67$, the quasi-flat bands are at the Fermi surface for $\nu=5/16$ or $\nu=11/16$, i.e., the lowest five or eleven bands are filled, see Fig.~\ref{main_fig}(f). In this way, the parameter $\alpha$ can be used to control the density of states at the Fermi surface, which will later be shown to have a strong impact on the superconducting behavior. We have defined the filling $\nu$ such that $\nu=1$ corresponds to a lattice which is filled by 4 fermions per site. We consider only the uniform value of filling (not distinguishing the filling at each synthetic level). In the limiting case of $\alpha=1$, twelve bands become fully degenerated at $E=0$, as can be seen from Fig.~\ref{main_fig}(b). 


In order to investigate the superconductivity, we consider attractive collisional Hubbard-type interactions between the atoms, i.e., local interactions in the physical lattice. We assume that the internal state of the atoms is not changed during the collision. In the most general form the interaction Hamiltonian then reads
\begin{equation}
    H_{\rm int}=-\sum_{\textbf{r},m,m',\sigma,\sigma'}U^{m,m'}_{\sigma,\sigma'} n_{m,\sigma}(\textbf{r}) n_{m',\sigma'}(\textbf{r}),
    \label{ham_int}
\end{equation}
where $n_{m,\sigma}({\bf r})=a_{m,\sigma}^\dag({\bf r}) a_{m,\sigma}({\bf r})$ is the density operator of a fermion in the $\{m,\sigma\}$ state.  $U^{m,m'}_{\sigma,\sigma'}$ denotes the (non-negative) interaction strength between atoms in levels $\{m,\sigma\}$ and $\{m',\sigma'\}$ and the negative sign indicates that we consider these interactions to be attractive. In general, $U^{m,m'}_{\sigma,\sigma'}$ describes 16 different kinds of interactions. Out of these, the Pauli principle excludes all diagonal interactions, i.e.,  $U^{m,m}_{\sigma,\sigma}=0$. We are left with possibly twelve different non-zero interaction processes. Because of symmetry arguments, these can be further grouped into three interaction types, as illustrated in Fig.~\ref{main_fig}(c):
The first type are the intralayer interactions within a synthetic layer
\begin{equation}
    U_1\equiv U^{+,+}_{\uparrow,\downarrow}=U^{-,-}_{\uparrow,\downarrow}=U^{+,+}_{\downarrow,\uparrow}=U^{-,-}_{\downarrow,\uparrow}.
\end{equation}
The second type groups the inter-layer interactions between particles of opposite spin,
\begin{equation}
    U_2\equiv
    U^{+,-}_{\uparrow,\downarrow}=U^{-,+}_{\uparrow,\downarrow}=U^{+,-}_{\downarrow,\uparrow}=U^{-,+}_{\downarrow,\uparrow}.
\end{equation}
The third groups contains the interlayer interactions of particles with equal spin
\begin{equation}
    U_3\equiv
    U^{+,-}_{\uparrow,\uparrow}=U^{-,+}_{\uparrow,\uparrow}=U^{+,-}_{\downarrow,\downarrow}=U^{-,+}_{\downarrow,\downarrow}.
\end{equation}
From the point of view of a realization with alkali-earth atoms, the case of $U_1=U_2=U_3$ is the most natural/realistic one. Interactions with nearly SU(N) symmetry are exhibited, for instance, between the internal states obtained from the nuclear spin manifolds ($I=5/2$ and $I=9/2$, respectively) for the fermionic isotopes $^{87}{\rm Sr}$ and $^{173}{\rm Yb}$, see Ref. \cite{cazalilla14}. We also discuss the cases in which $U_1$ (or $U_2$) become the dominant interactions, which is particularly relevant from the point of view of bilayer interpretation. 

\section{\label{MF}Mean-field decoupling}
We apply a Hartree-Fock-Bogoliubov-de Gennes mean-field approach \cite{Koch2016,goodman1979} to tackle the many-body Hamiltonian $H_\mathrm{kin}+H_\mathrm{int}$ (see Eqs.~\eqref{eq:Ht} and \eqref{ham_int}). As we consider attractive interactions, we  focus only on  pairing fields in the mean-field decomposition. Each on-site quadratic attractive term in Eq.~\eqref{ham_int} is thus decoupled  as:
\begin{equation}
\begin{split}
    a_{m,\sigma}^\dag a_{m,\sigma}a_{m',\sigma'}^\dag a_{m',\sigma'}
    &\approx
    \langle a_{m,\sigma}^\dag a_{m',\sigma'}^\dag\rangle a_{m',\sigma'}a_{m,\sigma}\\
    &+  a_{m,\sigma}^\dag a_{m',\sigma'}^\dag\langle  a_{m',\sigma'}a_{m,\sigma}\rangle\\
    &- \langle a_{m,\sigma}^\dag a_{m',\sigma'}^\dag\rangle\langle a_{m',\sigma'}a_{m,\sigma}\rangle,
\end{split}
\label{eq_mfdecouple}
\end{equation}
where $\langle\cdot\rangle$ denotes the average. The last term is a constant shift affecting the grand thermodynamic potential and is important for obtaining the self consistent equations for the order parameters via minimization of the thermodynamic potential or for assessing thermodynamic stability of different solutions. We do not display this term in the following. Combining Eqs.~\eqref{ham_int}-\eqref{eq_mfdecouple} and assuming symmetry between the layers and spins, let us explicitly write down the pairing Hamiltonian:
\begin{align}
\begin{split}
    H_{\rm P} &=  \Delta_1 ( a_{+,\uparrow}^\dagger a_{+,\downarrow}^\dagger  + a_{-,\uparrow}^\dagger a_{-,\downarrow}^\dagger )\\
    &+ \Delta_2 ( a_{+,\uparrow}^\dagger a_{-,\downarrow}^\dagger +  a_{+,\downarrow}^\dagger a_{-,\uparrow}^\dagger )\\ 
    &+ \Delta_3 ( a_{+,\uparrow}^\dagger a_{-,\uparrow}^\dagger +
    a_{+,\downarrow}^\dagger a_{-,\downarrow}^\dagger
    )
    + {\rm h.c.}
\end{split}
\end{align}
Here, we have defined the following superconducting order parameters
\begin{align}
    \Delta_1 &\equiv U_1\langle a_{+,\uparrow} a_{+,\downarrow} \rangle =  U_1\langle a_{-,\uparrow} a_{-,\downarrow} \rangle, \\
    \Delta_2 &\equiv U_2\langle a_{+,\uparrow} a_{-,\downarrow} \rangle =  U_2\langle a_{-,\uparrow} a_{+,\downarrow} \rangle, \\
    \Delta_3 &\equiv U_3\langle a_{+,\uparrow} a_{-,\uparrow} \rangle =  U_3\langle a_{+,\downarrow} a_{-,\downarrow} \rangle.
\end{align}

The value of the order parameters can, in principle,  vary within each unit cell due to in-equivalence of the lattice sites and their surroundings caused by spatial modulation of the synthetic coupling, described in Eq.~\eqref{homega}. As has been shown in Fig.~\ref{main_fig}(a) and defined in Eq.~\eqref{eq_unitcell},
one can differentiate the sites in the unit cell into four types. While types $B$ and $C$ are taken into account separately due to their different value of synthetic coupling $\Omega({\bf r})$, the sites of A-type (black sites in Fig.\ref{main_fig}) are distinguished based on  geometric reasons and divided into "bridge" (AA) and "node" (AB) sites, depending on the amount of nearest A-type neighbours (two for AA, and four for AB). A visual representation of this scheme is also shown in Fig.~\ref{sun_gap}, where yellow sites represent "bridge" (AA) sites on panel (a) and "node" (AB) on panel (b). Therefore, distinguishing between the four different types of sites, we write the order parameter in the interaction channel $i=1,2,3$ for site of type $I\in\{AA,AB,B,C\}$. For the eight sites of a unit cell, we thus have three sets of order parameters defined below:
    \begin{equation}
    \vec{\Delta_i} = (\Delta_i^{AB}, \Delta_i^{AA}, \Delta_i^{AB}, \Delta_i^{AA}, \Delta_i^{AA}, \Delta_i^B, \Delta_i^{AA}, \Delta_i^C).        
    \end{equation}
It is also convenient to view the real-space fermionic operators $a_{m,\sigma}({\bf r}_j)$ as eight-component vectors, with each component representing one site in the unit cell,  and the $j$ representing the index of unit cell within the Moir\'{e} lattice. Then, we Fourier transform the operators to quasi-momentum space, via
\begin{equation}
a_{m,\sigma}({\bf r}_j) = \frac{1}{\sqrt{N_s}}\sum_k e^{-i{\bf k}{\bf r}_j}a_{m,\sigma}({\bf k}), 
\end{equation}
where $N_s$ is the number of unit cells in the lattice and $c_{m,\sigma}({\bf k})$ is the eight-dimensional field operator of a fermion with quasi-momentum ${\bf k}=(k_x,k_y)$.
The real-space hopping $H_t$ is diagonalized as $H_t= -t \sum_{m,\sigma} a_{m,\sigma}^\dag({\bf k}) H^m_t({\bf k}) a_{m,\sigma}({\bf k})$, with $H^m_t({\bf k})$ being a diagonal matrix representing the eight bands per layer index $m\in\{+,-\}$. Due to the symmetry between the synthetic layers, we have $H^+_t({\bf k})=H^-_t({\bf k})$. The interlayer tunneling is also diagonal in $\bf k$, but of the form  $a_{+,\sigma}^\dag({\bf k}) H_\Omega({\bf k}) a_{-,\sigma}({\bf k})$.
\\
\\
In order to present the full Hamiltonian, $H=H_{\rm kin}+H_{P}$, containing all order parameters of interest in the quadratic form, we define the following $8\times8$-dimensional Nambu spinor:
\\
\\
\begin{equation}
  \Psi_\textbf{k}^\dag = 
  \begin{pmatrix}
  a_{-,\uparrow}^\dag({\bf k})\\
  a_{-,\downarrow}({\bf -k})\\
  a _{+,\uparrow}^\dag({\bf k})\\
  a_{+,\downarrow}({\bf -k})\\
  a_{-,\uparrow}({\bf k})\\
  a_{-,\downarrow}^\dag({\bf -k})\\
  a_{+,\uparrow}({\bf k})\\
  a_{+,\downarrow}^\dag({\bf -k})\\
  \end{pmatrix}
\end{equation}
The mean field Hamiltonian in momentum space is of the Bogoliubov-de-Gennes (BdG) form and is given by
\begin{equation}
   H(\textbf{k}) = \Psi_\textbf{k}^\dagger H_{\rm BdG}(\textbf{k}) \Psi_{\textbf{k}}-\epsilon(\textbf{k}),
\end{equation}
where $\epsilon(\textbf{k})$ is a diagonal matrix which includes all constant values coming from the decoupling. Note that $H_{\rm BdG}({\bf k})$ is a $64\times64$ matrix as each component of the Nambu spinor is an 8-vector yielding eight bands. The structure of this matrix is constructed as follows:
\begin{equation}
  H_{\rm BdG}(\textbf{k})= \begin{pmatrix} H_{F}({\bf k}) &  J_2\otimes\mathbf{1}\otimes {\rm diag}(\vec{\Delta_3}) \\ J_2\otimes\mathbf{1}\otimes {\rm diag}(\vec{\Delta_3}) &  H_F^{*}({\bf k})\\
  \end{pmatrix},
  \label{eq_BdG}
\end{equation}
where $\mathbf{1}$ is the two-dimensional unity matrix, $J_2=[[0,1],[1,0]]$ is the first Pauli matrix, and $\otimes$ denotes the tensor product.

The matrix on the diagonal block has the following structure
\begin{equation}
  H_{F}(\textbf{k})= \begin{pmatrix} H^{m}({\bf k}) &  H_{R}({\bf k}) \\ H_{R}({\bf k}) &  H^{m}({\bf k})\\
  \end{pmatrix},
\end{equation}
where the four $16\times16$ blocks are defined as 
\begin{widetext}
\begin{equation}
   H^{m}(\textbf{k})= \begin{pmatrix} H_{t\uparrow}^{m}(\textbf{k})-\mu-\frac{n}{3}(\frac{U_1}{2}+\frac{U_2}{2}+\frac{U_3}{2}) &  {\rm diag}(\vec{\Delta_1}) \\ {\rm diag}(\vec{\Delta_1}) &  -H_{t\downarrow}^m(-\textbf{k})+\mu+\frac{n}{3}(\frac{U_1}{2}+\frac{U_2}{2}+\frac{U_3}{2})\\
   \end{pmatrix},
\end{equation}
\end{widetext}
and
\begin{equation}
  H_{R}(\textbf{k})= \begin{pmatrix} H_\Omega(\textbf{k}) & {\rm diag}(\vec{\Delta_2}) \\ {\rm diag}(\vec{\Delta_2}) & -H_\Omega(-\textbf{k})
  \end{pmatrix}.
\end{equation}
The quadratic matrix $H(\textbf{k})$ depends on the unknown superconducting order parameters $\vec{\Delta_i}$, which we determine self-consistently by diagonalizing $H(\textbf{k})$ using random initial guesses of $\vec{\Delta_i}$, and subsequently updating the order parameters by the ones obtained from the diagonalization until convergence is attained. We check that this procedure leads to the same order parameters for different initial guesses, or, if this is not the case,  we choose the solution with the lowest grand thermodynamic potential energy.

\section{\label{results}Results}
In this section we present the original results of the pairing correlations in the system within the framework described above.  While we first focus on the case of fully symmetric interactions that naturally arise in the context of the experimental proposal presented in \cite{Salamon_2020}, namely $U_1$=$U_2$=$U_3$ (see Fig.~\ref{main_fig}(c)),  where all internal degrees of freedom of each atom are coupled to the each others with the same strength, we also study the effects of SU(4) symmetry breaking by considering the relative alteration between interaction channels $U_1$ and $U_2$ in the subsequent subsection. In this context, we set $U_3=0$ in all the subsequent calculations, which is justified because any pairing in the $U_3$ channel is strongly suppressed by the strong Raman coupling, $\Omega$. This coupling (interlayer hopping) energetically penalizes the state with two particles of equal pseudospin per site, as compared to the single-particle states formed by the antisymmetric superposition of the states with equal pseudospin and opposite pseudolayer degree of freedom.
\\

A very interesting phase diagram is found in the regime of weak interactions: superconductivity is exponentially suppressed in the symmetric case, i.e. near $U_1=U_2$, but a significant non-zero SC gap can be again amplified if the interactions are tuned to a sufficiently non-symmetric choice. We observe the narrowing of the weakly-superconducting wedge shaped region in the phase diagram with increasing interaction strength, as well as the coexistence of both the inter and intra-layer superconducting order parameters. Techniques to tune interactions, such as Feshbach resonance or magnetic/optical field gradients \cite{Sonderhouse2020,Martone14}, allow in experiments for such a selective choice of a dominant interaction channel. Increase of band flattening  via the tunable parameter $\alpha$ leads to the development of strong superconducting order for lower values of attractive interactions, in particular, as compared to the  standard ($\alpha=0$) bi-layer Fermi-Hubbard model. 

Numerical calculations were performed for a system of 2 $\times$ 256 sites forming a $2\times L \times L$ bilayer square lattice with $L=16$. The low temperature properties were studied by choosing the temperature to  $ k_B T=0.02t$ (or inverse temperature $\beta \approx 3 L$ for the finite size system). 


\subsection{Superconductivity in $SU(N)$ symmetric system}
\label{sec_gapBoost}

\begin{figure}
    \centering
    \includegraphics[width=1\linewidth]{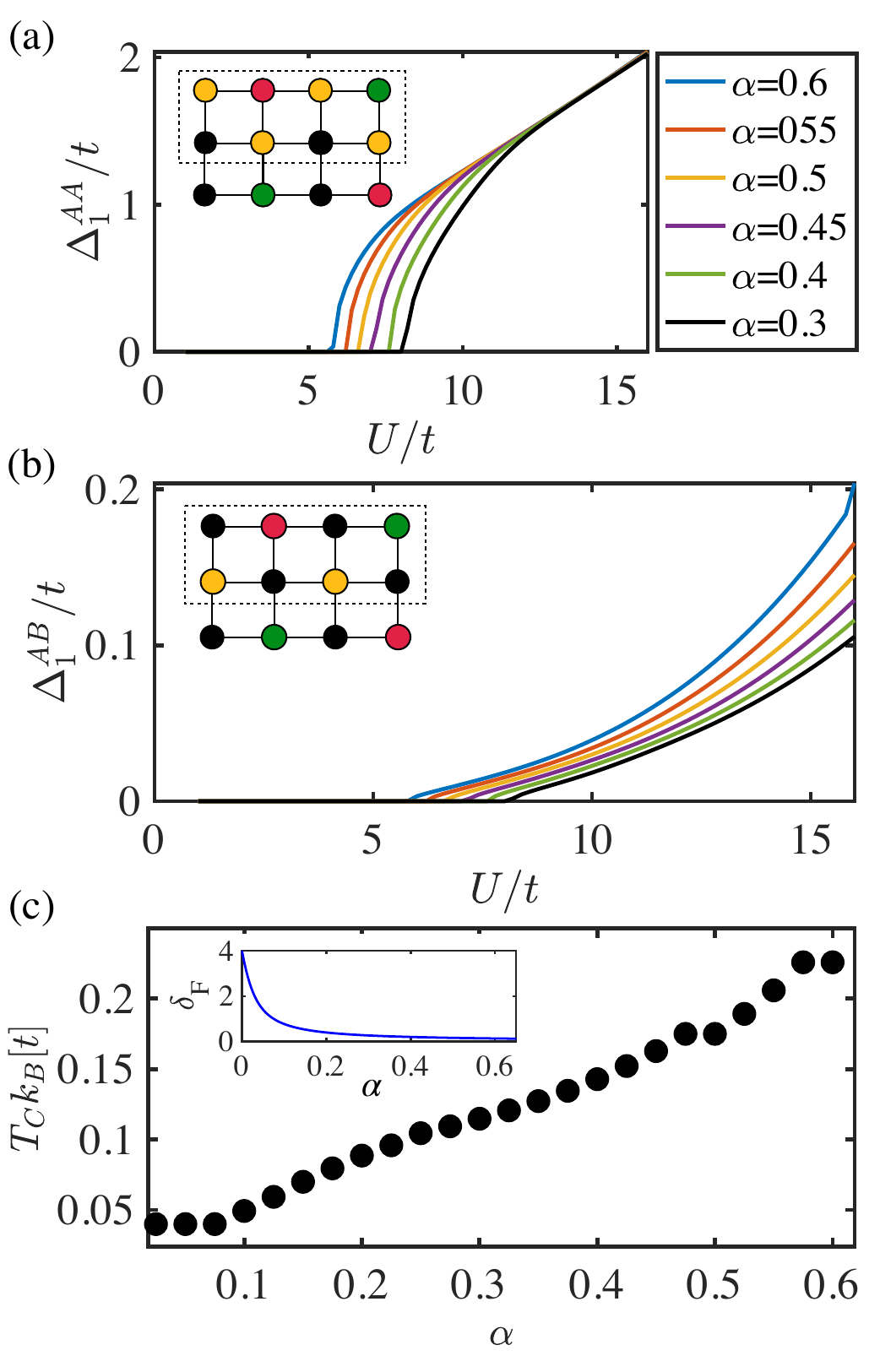}
    \caption{(a,b) Expansion of superconducting gaps $\Delta^{AA}_{1}$ and $\Delta^{AB}_{1}$ as a function of the $U=U_1=U_2\leq16t$ under different modulation values $\alpha$ resulting in specific cutoff value $U_c$ ($\Delta_{2}$ has an identical behavior). (c) Critical temperature dependence on the modulation strength $\alpha$ at $U_{1,2,3}=16t$. Inset plot represents the energetic width $\delta_F$ of the quasi-flat bands as a function of the modulation parameter $\alpha$.}
    \label{sun_gap}
\end{figure}

We set a strong interlayer (Raman) coupling  $\Omega_0/t=100$ in order to focus on effects in the quasi-flat band regime of our system. We investigate the influence of two main parameters, namely the modulation amplitude $\alpha$ and interaction strength $U$ on the SC characteristics. 
\\
\\
The parameter $\alpha$ controls the relative strength of the  spatially dependent part of the synthetic hopping $\Omega(\mathbf{r})$ in eq.~\eqref{omega_form}. The increase of $\alpha$ primarily results in flattening in the dispersion of the set of quasiflat bands of our interest. Up to a value of $\alpha\approx 0.67$, the flattened bands lie exactly at Fermi energy of the system for a filling $\nu=1/4$ (that is, one fermion per physical site). As shown on Fig.~\ref{main_fig}(b), the critical value of $\alpha\approx 0.67$ causes  band flipping, which lifts the Fermi energy of the quarter-filled system away from quasi-flat band (see  Fig.~\ref{main_fig}(d) and (f)). We focus on the case with $\nu=1/4$ and $0\leq\alpha\leq0.67$. A large superconducting gap opens in the SU(4) symmetric system, for interaction strengths, higher than a certain, $\alpha$-dependent cut-off interaction amplitude, namely $U>U_C(\alpha)$. We note that $U_C(\alpha)$ decreases if one considers lower temperatures. 
Enhancing the density of states at the Fermi energy by increasing the band flattening with increasing $\alpha$ is expected to lead to larger stability of the superconducting phase and therefore lower threshold values $U_C(\alpha)$.Indeed, such behaviour is markedly seen in Fig.~\ref{sun_gap}(a,b). For an example value of $\alpha=0.6$, the superconducting state appears above $U \sim 6.2t$. Similar effect has been observed for the critical temperature of the system with respect to modulation $\alpha$. Fig.~\ref{sun_gap}(c) depicts the growth of the critical temperature with band-flattening.  We have limited the range of $\alpha$ in the figure such that it covers only the scenario with Fermi energy matching the energy of quasi-flat bands. Therefore, by tuning the modulation parameter $\alpha$ one can reach superconducting state at lower interaction values as compared to the uniform inter-layer hopping scheme. Moreover, resulting difference in critical temperature between highly modulated system at  $\alpha=0.6$ and un-modulated one is of one order of magnitude. In other words, the superconductivity near $U_C$ is truly induced by the synthetic twist.

\subsection{$SU(4)$ to $SU(2)\times SU(2)$ symmetry breaking}
\label{sec_SU_N_breaking}

\begin{figure*}
\makebox[\textwidth][c]{\includegraphics[width=0.9\textwidth]{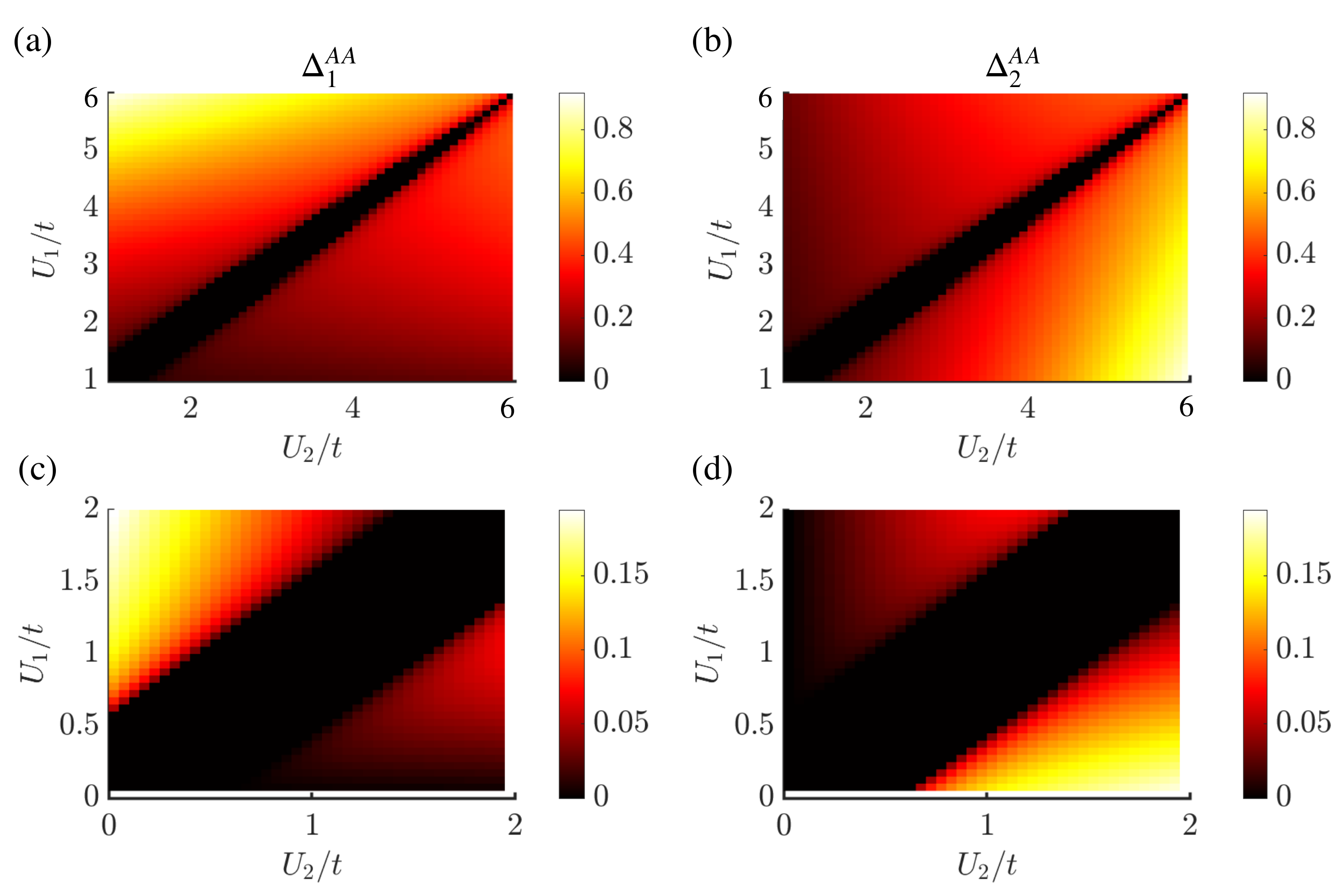}}%
\caption{Superconducting order parameter $\Delta_1^{AA}$ (panel a and c) and $\Delta_2^{AA}$ (panel b and d) as functions of relative amplitude of $U_1$ and $U_2$ interactions at $\alpha=0.67$. Existence of SC phase in this $SU(4)$ symmetric system is not present until $U=U_1=U_2$ reaches a cut-off value $U_c$, at which the non-SC valley narrows to zero. The width of the valley as well as the critical interaction strength depends on the modulation parameter $\alpha$ (as shown on the Fig.~\ref{sun_gap}). Panels (c) and (d) represent the regime of small interaction strengths that are not covered on the panels (a) and (b) due to limited colormap resolution. Sites of (AB) type reveal qualitatively identical behavior, however, with much smaller pairing amplitude (See Fig.\ref{fig_maps}(b))}.
\label{gap_closing}
\end{figure*}

The SU(N=4) symmetry (spin and magnetic levels) of the system resulting from the internal structure of the atoms forbid the free tuning of the strength of each interaction type individually without an external fields applied. As shown in \cite{Stellmer11,Sonderhouse2020}, one can tune the strength of interaction types by applying external state-dependent force that effectively separates the $m_F$ manifold of the ground state. For atoms with two valence electrons, such as $^{87}$Sr, this technique, also termed as, optical Stern-Gerlach (OSG) has been already successfully applied experimentally \cite{Stellmer11,Sonderhouse2020}. Following the scheme of the system proposed in \cite{Salamon_2020,Sala2}, we propose to use OSG to modify the energy gaps between specific $m_F$ states, thereby tuning the interaction strength of desired type. In this paragraph, we study the effects on the SC properties due to a modification of the interaction strength in such fashion. We note that the extreme scenario of $U_1 \gg U_2$ corresponds to standard spin-spin onsite only interactions widely explored in Fermi-Hubbard model. Whereas the SU(N)-symmetric scenario requires a threshold value of the coupling $U$ for the system to exhibit SC, the symmetry-broken scenario allows us to observe SC pairing even for $U_1,U_2<U_C$ (with one of them possibly even being zero). Interestingly, a phase where SC is dominated by $U_1$ is separated from a $U_2$ dominated SC phase through an intermediate phase in which SC is absent, leading to the interesting re-entrance phenomena, when one of the interaction parameters is tuned.
\\
\begin{figure}[h]
\includegraphics[width=1\linewidth]{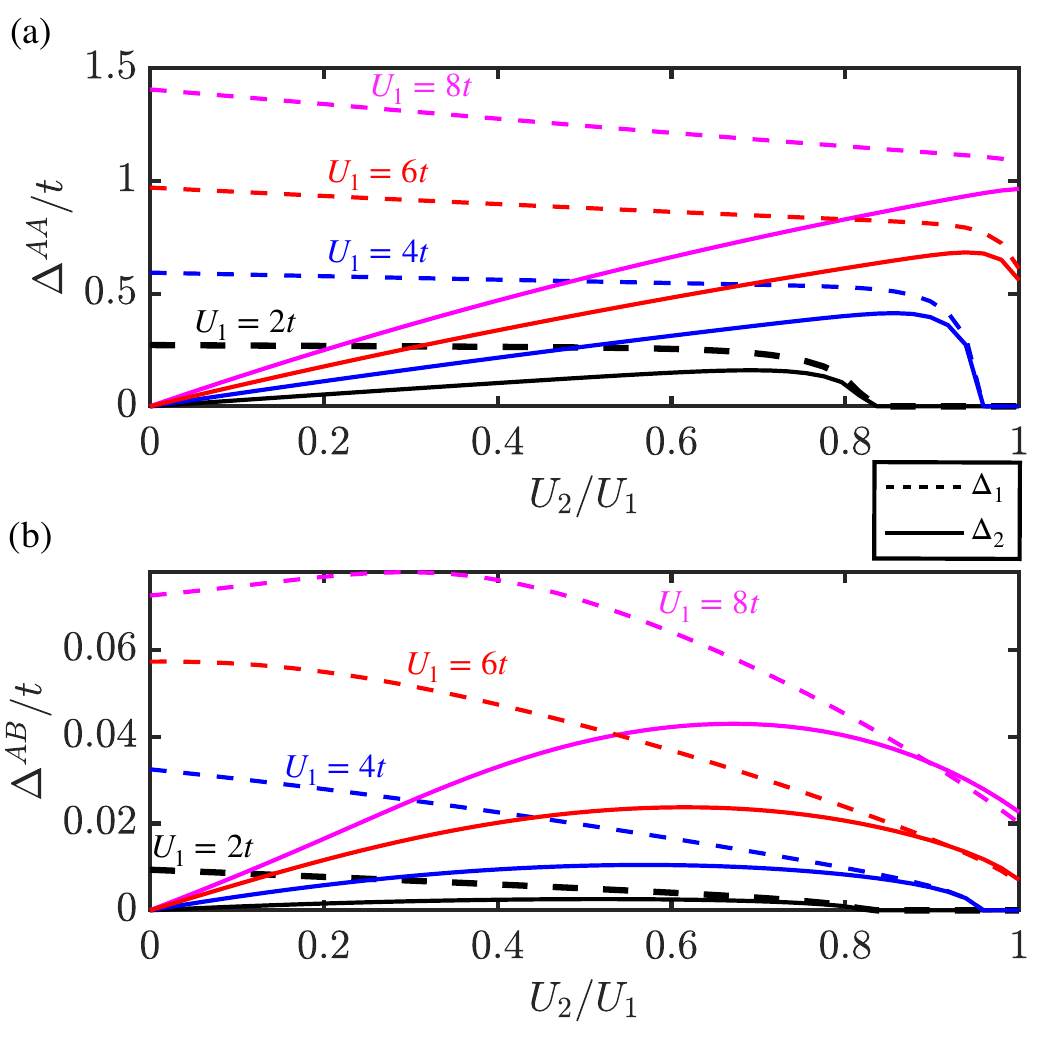}
\caption{ $\Delta_{AA,AB}$ (dashed) and $\Delta_2$ (solid) as functions of $U_2$ (in units of $U_1$) for four different values of $U_1$ at modulation strength $\alpha=0.67$. One can observe the shift of the SC gap decay towards $U_2/U_1=1$ with the increase of interaction strength. For $U_1=6t$, the decay of SC gap occurs at $U_2/U_1=1$ marking the $U_C$ for this particular value of $\alpha$.}
\label{fig_maps}
\end{figure}

The different phases are seen in Fig.~\ref{gap_closing} where the SC gaps $\Delta^{AA,AB}_1$ are plotted as a function of the interaction strengths $U_1$ and $U_2$, in the interval $[1,6]$ in panels (a) and (b), and in the interval $[0,2]$ in panels (c) and (d). Here we have chosen $\alpha=0.$, and the corresponding $U_C$ from the SU(4) symmetric system is $U_C\approx6$. Hence, the shown regime is below $U_C$ everywhere, and accordingly, the system does not exhibit SC along the line $U_1=U_2$. It is seen that this non-superconducting regime, plotted in black, has a finite width, which narrows as $U_1$ and $U_2$ approach towards $U_C$. The width and the rate of the narrowing depends on the modulation parameter $\alpha$, as indicated in the Fig.~\ref{fig_maps}.
Nevertheless, relatively small deviations from the symmetric interaction are already sufficient to open a SC gap. This can be understood in the following way: the SC pairing of each interaction type compete with each other, but breaking the symmetry favors one interaction type with respect to the other, and therefore facilitates the pairing in this channel.
\begin{figure}[ht]
\includegraphics[width=1\linewidth]{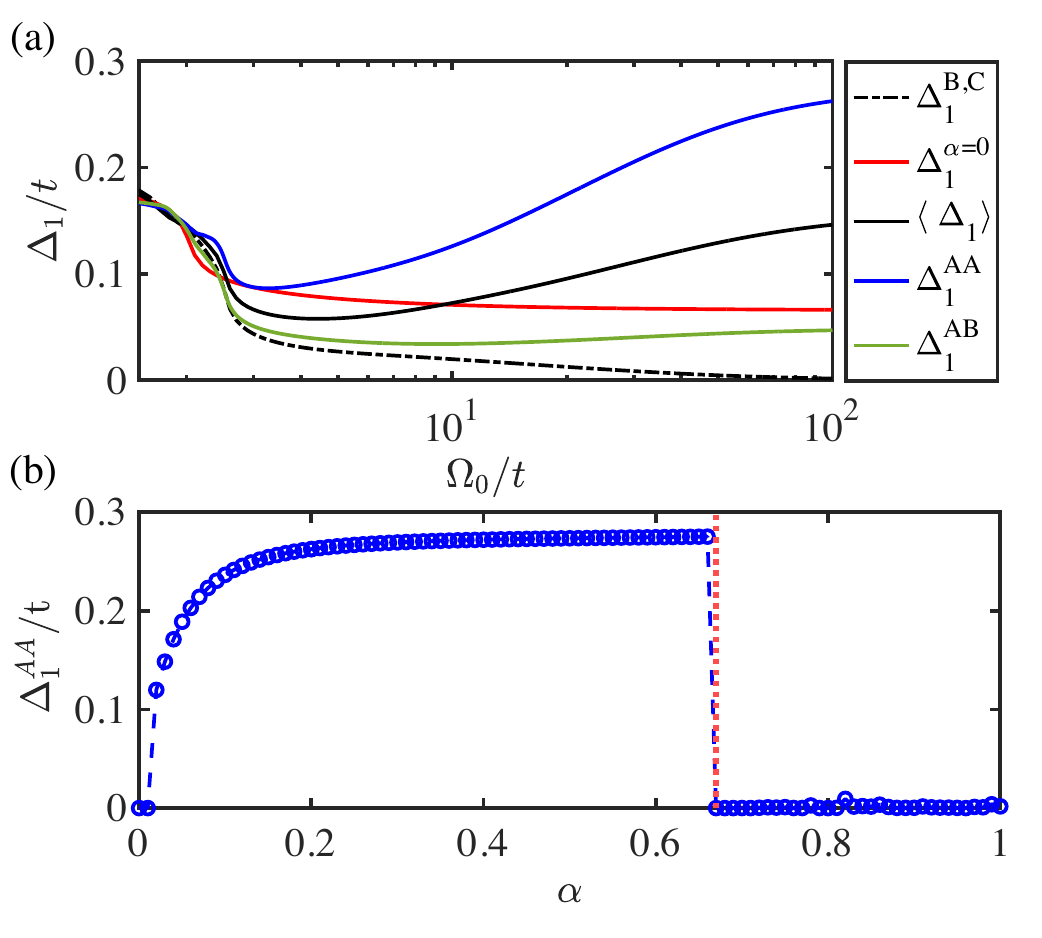}
\caption{Superconducting gap in the system of $U_{1}=2t$ and $U_{1} \gg U_{2}$. (a) The order parameter $\Delta_{1}$ associated with interaction type $U_{1}$ as a function of the inter-layer hopping strength $\Omega_0$ at $\alpha=0.2$. We have set $U_{2}=0$ as its presence weakens the order parameter $\Delta_{1}$, as shown on Fig.~\ref{fig_maps}(b). The green line represents the situation where the $\Omega(\textbf{r})$ is homogeneous, i.e. $\alpha=0$. The separate plot for $\Delta_C$ has been omitted since it's behaviour is identical to $\Delta_B$, which has been plotted. (b) Dependence of the order parameter $\Delta_1^{AA}$ on the modulation strength $\alpha$ for $\Omega_0=100t$. We have omitted the plot of $\Delta_1^{AB}$ due to its negligible amplitude. Sudden drop of pairing (marked by dashed gry line) occurs around critical value of $\alpha\approx2/3$ at which the Fermi level does not reside at quasi-flat bands any more. Such situation has been shown on Fig.~\ref{main_fig}(f).}
\label{alpha_dep_tot}
\end{figure}
Although one of the two different pairing channels becomes dominant, there is still coexistence of the SC gaps corresponding to the two channels, $U_1$ and $U_2$, for most parameter choices (unless we are in the non-SC regime, or one of the interaction parameters is zero).
However, monotonicity of the size of the gap with respect to the interaction strength is not obvious. We present a closer insight to this phenomenon through the details presented in Fig.~\ref{fig_maps}. 
It depicts the dependence of the SC order parameters on $U_2$ for four different values of $U_1$ at $\alpha=0.67$. The decay of $\Delta^{AA}_{1}$ and $\Delta^{AA}_{2}$ as $U_2$ approaches $U_1=U_2$ is the beginning of the zero-gap valley. With increasing value of $U_2$ the corresponding gap $\Delta^{AA}_2$ opens and keeps increasing until reaching its maximum value. In parallel, SC order parameter corresponding to $U_1$ constantly decreases.

We now consider the cases of $U_{1} \gg U_{2}$ which can be obtained experimentally with help of OSG techniques. We would like to note that the same results have been obtained for the opposite case, i.e. $U_{2} \gg U_{1}$. This scenario qualitatively agrees with the standard bilayer Fermi-Hubbard model with only in-plane interactions typically considered as good approximations to describe various phenomena in condensed matter physics. Note, that in this paper we are not aiming at direct comparison with real solid state systems. In contrast, our focus here is on a non-standard choice of parameters ($\Omega \gg t$) as it gives us access to study the effect of a flat band structure even in relatively small Moir\'e supercells. Such regime of parameters is accessible in cold atomic systems. We choose this regime in order to obtain a maximum possible value of the gap for a given amplitude of $U$. As it has been shown on the Fig.~\ref{fig_maps}, the widest gap appears for highly unequal interaction values, i.e. \textbf{$U_{1} \gg U_{2}$}. We begin by investigating the dependence on the synthetic hopping amplitude, $\Omega_0$. In its absence the system consists of two uncoupled layers of square lattices. At finite $\Omega_0$, we can flatten the bands through the spatial modulation provided by $\alpha$, or realize the standard bi-layer model, i.e. $\alpha=0$. Panel (a) of Fig.~\ref{alpha_dep_tot} depicts both of these scenarios. In particular, we have plotted separately the gap for A-type and (B,C) sites in $\alpha$-modulated case, as well as the mean value of the gap averaged over all sites of the unit cell. The green line represents the size of the gap for a standard bilayer model, that is, with $\alpha=0$. Interestingly, with a small separation of the spectrum caused by inter-layer hopping, SC gaps drop. However, after a full separation of the bands into positive and negative branches, the system the size of the gap of the quasi-flat band system starts to grow, in contrast to the SC order parameter of the standard FH system.  Panel (b) of Fig.~\ref{alpha_dep_tot} represent the dependence of the order parameter $\Delta_{1}$ on the modulation parameter $\alpha$. The plot depicts the situation where $\Omega_0/t=100$ and therefore $\Delta^{B,C}_{1}$ can be neglected due to their vanishing values. The amplitudes of $\Delta_{1}$ in panel (b) have been obtained for the fixed filling $n=1/4$. 
Summarizing, modulation of the inter-layer hopping leads to enhanced SC order parameter with respect to a non-modulated one at sufficiently high $\Omega_0$. This effect is a result of band flattening and therefore disappears once the modulation $\alpha$ crosses the critical value or the Fermi energy does not match the energy of quasi-flat bands. Qualitatively similar results also follow for larger finite $U_2/U_1$.

\section{\label{Conclusions}Conclusions}
In this paper, we have used Bogoliubov-deGennes theory to study attractive interactions of the synthetic bi-layer square lattice system. The studied model goes beyond the thoroughly explored bi-layer Hubbard models and tackles correlated phases in quasi-flat band systems emerging from the periodic modulation of interlayer hopping. We have taken into account all possible density-density on-site interaction channels and considered properties across a wide range of experimentally accessbile interaction strengths. Our system has a small Moir{\'e} unit cell for which flat band induced effects occur for large interlayer hopping strengths. We note that similar small Moire unit cells generated at large twist angles in physical bilayers would also require rather large interlayer hopping to isolate the flat band regime. This could in principle be achieved by applying strong strain or pressure in the direction perpendicular to the plane of the layers in materials. However, it is rather more easily achieved in our synthetic system where the interlayer hopping is controlled simply by the intensity of a Raman laser coupling the internal levels that play the role of the layer degrees of freedom.\\ 

First, we have focused on the natural case of equal interaction amplitudes and observed strong dependence between the inter-layer modulation parameter $\alpha$ and minimal interaction strength $U_c$ required to reach SC pairing at a fixed low temperature. Observed results confirm the following: {\emph{(i) Flattening of the bands in the vicinity of Fermi energy leads to opening of the SC gap at much lower interaction amplitudes, when compared to uniform coupling systems, and (ii) Band flattening causes the critical temperature scale to significantly increase in these novel synthetic-twist-induced  latices with magic configurations and thus superconductivity (paired neutral fermion superfluidity) may be potentially observable in state-of-the-art cold gas experiments.}}\\ 

Further results are obtained by altering relative interaction amplitudes in the system: { \emph{(iii) The resulting phase diagram revealed  a valley around $U_1=U_2$ with strongly suppresssed superconducting correlations. The width of this valley narrows with the growing amplitude of interactions to finally completely vanish at $U_c$ specific for each value of $\alpha$.}}  Similar behaviour has been observed in bi-layer Hubbard square lattices with only one correlated layer \cite{Zujev_2014}. There, the apparent re-entrance of the SC gap was a result of increasing inter-layer hopping. Here, however, we observe similar effects as a function of inter-layer interactions for large inter-layer hopping.\\
\\

While the results discussed in this paper help to form a general understanding of the effects of band flattening on superconductivity in the synthetically twisted materials, further studies can be be pursued in future, particularly, in context of  topologically nontrivial bands with relatively weak dispersion, which can be obtained via more involved, but experimentally viable, means e.g., via imaginary next-to-nearest neighbor tunnelings driving the system into a quantum anomalous Hall phase \cite{Salamon_2021}.  In such cases, Wannier functions with algebraically decaying tails may originate from nonzero Chern number \cite{Peotta_2015}.  Novel understandings  of the correlated phenomenon could then be obtained via incorporation of new mechanisms, such as  correlated tunnelings in extended Hubbard bi-layer systems.

\section{\label{acknos}Acknowledgements}
The authors would like to thank Utso Bhattacharya and Leticia Tarruell for all the fruitful discussions and insightful comments. We acknowledge support from: ERC AdG NOQIA; Agencia Estatal de Investigación (R\&D project CEX2019-000910-S, funded by MCIN/ AEI/10.13039/501100011033, Plan National FIDEUA PID2019-106901GB-I00, FPI, QUANTERA MAQS PCI2019-111828-2, Proyectos de I+D+I “Retos Colaboración” QUSPIN RTC2019-007196-7);  Fundació Cellex; Fundació Mir-Puig; Generalitat de Catalunya through the European Social Fund FEDER and CERCA program (AGAUR Grant No. 2017 SGR 134, QuantumCAT \ U16-011424, co-funded by ERDF Operational Program of Catalonia 2014-2020); EU Horizon 2020 FET-OPEN OPTOlogic (Grant No 899794); National Science Centre, Poland (Symfonia Grant No. 2016/20/W/ST4/00314); European Union’s Horizon 2020 research and innovation programme under the Marie-Skłodowska-Curie grant agreement No 101029393 (STREDCH) and No 847648  (“La Caixa” Junior Leaders fellowships ID100010434: LCF/BQ/PI19/11690013, LCF/BQ/PI20/11760031,  LCF/BQ/PR20/11770012, LCF/BQ/PR21/11840013). RWC acknowledges support from the Polish National Science Centre (NCN) under the Maestro Grant No. DEC-2019/34/A/ST2/00081. DR acknowledges support from Science and Engeenering Research Board (SERB), Department of Science and Technology (DST), under the sanction No. SRG/2021/002316-G. 

We would like to warmly dedicate this paper to the memory of Roman Micnas. 

\section{Data availability} The data that support the findings of this study are available from the corresponding author, T.S, upon reasonable request.
\\
\\
\\
\\

\newpage

\bibliographystyle{apsrev4-1}

%


\end{document}